  \documentstyle[prb,aps,twocolumn,floats,epsfig,twoside]{revtex}
  \input entete.tex
\def \la{\left(}
\def \ra{\right)}
\def \lb{\left[}              %
\def \rb{\right]}             

\def \ld{\left\langle}
\def \rd{\right\rangle}

\def \cH{{\cal H}}

\def \cO{{\cal O}}

\let\a=\alpha
\let\b=\beta
\let\d=\delta

\let\g=\gamma
\let\G=\Gamma
\let\k=\kappa
\let\l=\lambda
\let\L=\Lambda

\let\n=\nu
\let\r=\rho

\let\s=\sigma

\def \vec #1{\mbox{\boldmath ${#1}$}}

\def \be{\begin{equation}}
\def \ee{\end{equation}}
\def \bea{\begin{eqnarray}}
\def \eea{\end{eqnarray}}
\begin{document}
\draft
\twocolumn[\hsize\textwidth\columnwidth\hsize\csname@twocolumnfalse\endcsname

\title{Critical behaviour of the 2$\vec d$ spin diluted 
       Ising model via the equilibrium ensemble approach}

\author{Giorgio Mazzeo\cite{emazz} and Reimer K\"uhn\cite{eku}}

\address{Institut f\"ur Theoretische Physik, Universit\"at Heidelberg, 
         Philosophenweg 19, 69120 Heidelberg, Germany}

\date{June 4, 1999}
\maketitle
\begin{abstract}
The equilibrium ensemble approach to disordered systems is used
to investigate the critical behaviour of the two dimensional Ising model 
in presence of quenched random site dilution. The numerical transfer matrix 
technique in semi-infinite strips of finite width, together with 
phenomenological renormalization and conformal invariance, is particularly 
suited to put the equilibrium ensemble approach to work. A new method to 
extract with great precision the critical temperature of the model 
is proposed and applied. 
A more systematic 
finite-size scaling analysis 
than in previous numerical studies has been performed. A parallel 
investigation, along the lines of the two main scenarios currently under 
discussion, namely the logarithmic corrections scenario (with critical 
exponents fixed in the Ising universality class) versus the weak universality 
scenario (critical exponents varying with the degree of disorder), is 
carried out. In interpreting our data, maximum care is constantly taken to 
be open in both directions. A critical discussion shows that, still, an 
unambiguous discrimination between the two scenarios is not possible on the 
basis of the available finite size data.

\end{abstract}
%
%

]



\section{Introduction}

Recent years have witnessed renewed efforts in the statistical physics
community to understand phase transitions in simple disordered classical 
spin systems. As a matter of fact, these efforts have produced conflicting 
statements concerning the effects of disorder on critical phenomena, covering 
almost the complete spectrum of conceivable alternatives.

This holds in particular, but not exclusively, for two-dimensional (2$d$) 
disordered ferromagnetic (i.e., unfrustrated) Ising models. These models 
have been widely studied, both because the corresponding pure system is 
well understood and because they constitute a marginal case in the Harris 
criterion \cite{Har74} which assesses whether disorder constitutes a relevant 
or irrelevant perturbation for the critical behaviour of the pure system. 
For models of this type, the discussion currently appears to narrow down 
on two conflicting scenarios, namely the {\em logarithmic corrections\/}
\cite{DD83,Sha84,Shan87,Lu88,De+87,An+90,TalShu94,Sel+94,Aa+96,sldq97,%
Aa+97,St+97,Ball+97,Rod+98,Sel+98,Ple98} 
versus the {\em weak universality\/}
\cite{Ku87a,Fae92,KiPa94,Ku94} 
scenario, though a broader spectrum of 
alternatives had been discussed earlier \cite{Heu91,SouBM92,Zie90} 
(the interested reader will find a comprehensive report on the literature
up to approximately 1982 in an early review by Stinchcombe \cite{Sti83}).
We will describe these scenarios and discuss these (and related) results
in greater detail later on.

The object of our study is the randomly spin diluted 2$d$ Ising model. 
Our investigation is based on two main ingredients. First, we use 
the equilibrium
ensemble approach to disordered systems \cite{Mo64,SoWa79,Ku87a,Ku94,Ku96} to 
map the quenched system onto an equivalent thermodynamic equilibrium system in 
an enlarged phase space. Performing this mapping exactly would be tantamount 
to providing an exact solution to the original problem, which is clearly 
infeasible. So a scheme of approximations based on a moment matching idea is 
invoked. 
Second,
we resort to conventional transfer matrix (TM) techniques 
to implement the method in practice on finite-width strips, 
analyzing finite-size results in the line of 
Nightingale's phenomenological renormalization group scheme. \cite{Ni76,Ni79}

The purpose of the present paper is to provide details of our TM
study, \cite{Ku87a,Ku94} as well as to include new material and to describe 
significant advances
in the understanding of 
finite-size scaling (FSS) signatures in the presence of logarithmic
corrections, which together have allowed 
us to boost the accuracy of our results considerably
and to obtain a sounder appreciation of the subtleties 
that may emerge in the interpretation of the data.

The main and unexpected finding in Ref.\ \onlinecite{Ku94}
has been a continuous variation of 
the critical exponents $\a$, $\b$, $\g$ and $\n$ with the spin density $\r$ in 
a manner which was observed to comply with the idea of weak
universality. \cite{Suz74} That is, the exponent $\eta$ describing the decay of 
critical correlations, and the magnetic exponent $\d$, as well as the ratios 
$\b/\n$ and $\g/\n$ were found to be {\it independent\/} of $\r$. The results 
were obtained by extrapolations of FSS data based on 
rather moderate strip-widths, 
and they were in complete 
quantitative and qualitative agreement with those of a Monte Carlo study 
by Kim and Patrascioiu. \cite{KiPa94} 
These results are in conflict with those supporting the logarithmic 
corrections scenario, where 
modifications of the relevant thermodynamical quantities at criticality
appear through logarithmic terms in the reduced temperature, 
while the system is left in the same universality class (i.e.
with the same critical exponents) as that of the pure 2$d$ Ising
model. These findings are almost all concerned with the ferromagnetic
bond disordered situation, either theoretically, \cite{Sha84,Shan87,Lu88,Ple98}
or via numerical 
\cite{De+87,An+90,Sel+94,Aa+96,sldq97,Aa+97,St+97,Ball+97,Sel+98} 
and series expansion \cite{Rod+98} approaches,   
but the comparison with the spin diluted model is possible
supposing -- as it is generally believed to be the case  -- 
that the two systems are in the same universality class. \cite{Sti83,Ple98} 

In discussing our results, we have to address {\em two main issues\/}. The 
first is concerned with the reliability of our method, which is 
based on an approximate description of quenched disorder. 
The evidence we have been able to compile 
does give us strong confidence in the validity of our approach. This granted, 
we turn to the second issue: 
can -- or 
more precisely to what extent can -- our results provide evidence in favour 
of or against any of the conflicting scenarios so far advanced to describe 
the critical behaviour of 2$d$ disordered ferromagnetic Ising models? 
This is indeed a subtle question, and we devote almost two Sections 
(Secs.\ \ref{resu} and \ref{disc}) to discussing it. 
It turns out that a considerable portion 
of the available finite-size data from simulations or TM studies --
including those presented in our earlier study \cite{Ku87a,Ku94} 
as well as some new ones -- may perhaps {\em not\/} allow to decide 
with 
sufficient confidence 
between the two most serious candidates.
And we shall explain why. Larger system sizes would in 
any case be necessary in studies along those lines to permit taking sides. 

Finally, since one of the aims of our study is to discriminate
between these two contradicting scenarios, 
one of our main concerns is an open-minded attitude, 
perhaps not easily found in the literature, in the 
analysis of the numerical results: 
we take the greatest care not to select one of them 
a-priori and to fit our data to it, but we rather attempt to
be open in both  directions. 

The remainder of our paper is organized as follows. In Secs.\ \ref{syst}
and \ref{eea}, 
in order to set the scene and to fix our notation, we describe the model 
that is studied in the sequel and we briefly recall our methodical 
background, viz. the equilibrium ensemble approach to disordered systems
(EEA). 
While Sec.\ \ref{tran} presents 
some details related to the computational side of the problem, we reserve 
Sec.\ \ref{scen} for a brief review of the two above mentioned scenarios, 
and Sec.\ \ref{resu} for a presentation and discussion of our main results. 
We describe a new finite-size procedure to extract the critical 
temperature of the model as a function of the degree of dilution, 
which enables us to draw an accurate approximation of the phase diagram.
We then determine values for the critical exponents, for the central
charge and for the specific heat of 
our approximating systems. 
We end our paper with 
a critical report of the available finite-size data
obtained also by other authors, and with  
a comprehensive discussion both on the method and 
on the outcome of our investigation (Sec.\ \ref{disc}). An Appendix, mainly
based on renormalization group results by Cardy and Ludwig,\cite{Car86,Lud-Car} 
presents a systematic FSS analysis of thermodynamic quantities in the 
presence of logarithmic corrections. 
 

\section{Setting the Scene}
\subsection{The system under investigation}
\label{syst}

The 
object of our investigation is the 2$d$ randomly spin diluted Ising model,
described by the Hamiltonian 
\be
   \cH(\s|\k) = - J\sum_{\langle i j \rangle} k_i \s_i k_j \s_j 
                - H \sum_i k_i \s_i\ .
\label{modH}
\ee
Here, the first sum is over all nearest neighbour pairs 
$ \langle i j \rangle $ of, say, a 
square lattice $\L$, and the second over all sites $i$. The $\s_i$ denote 
Ising spins and the $k_i \in \{0,1\}$ are occupation numbers signifying whether 
in a disorder configuration $\k$ a site $i$ is occupied by a spin ($k_i=1$) or 
not ($k_i=0$). The $k_i$ are taken to be quenched random variables, i.e., they 
are fixed, and randomly chosen according to the probability 
\be
q(\k) = \prod_{i} \r^{k_i}(1-\r)^{1-k_i} \ ,
\ee
which simply requires each site to be occupied with probability $\r$, and
to be empty with probability $1-\r$. Thus $\r$ is the average density of spins 
in the system.

Our aim is to study the thermodynamics of the system, described by the 
(dimensionless) quenched free energy
\be
f_q = - N^{-1} \ld \ln Z_N(\k)\rd_{q}\ ,
\label{fqu}
\ee
i.e., the average of the system's free energy  over the distribution $q(\k)$ 
describing the statistics of the disorder configurations $\k$. Here $N$ denotes 
the system size, and $Z_N(\k)$ is the partition function of a system of size 
$|\L| = N$ at fixed disorder configuration $\k$. In particular, we would like 
to locate the  phase-transition line $T_c(\r)$ below which the system is 
known to develop spontaneous ferromagnetic order in the thermodynamic limit 
$N\to\infty$, and to study the critical behaviour of the system at and around 
this line. 
Whereas the thermodynamics of the 1$d$ system is known exactly, an exact 
solution of the 2$d$ model is currently way beyond our abilities. 
Approximation methods, which may, however, well produce exact results in 
certain limiting cases are thus called for.

\subsection{Methodical background: the equilibrium ensemble approach}
\label{eea} 

The EEA to disordered systems, as well as its application to 
the 2$d$ spin diluted Ising model, have been accurately described 
elsewhere. \cite{Ku96,Ku94}
Therefore we will not enter again into the details of the method, rather
just concisely recall the main ideas.  

The central idea is to treat the configurational degrees of freedom $k_i$ in a 
system with quenched disorder on the same footing as the dynamical variables 
proper -- the spins, in the case at hand. That is, the phase space is enlarged 
to include the occupation-number configurations $\k$, and an additional term
depending only on $\k$, the so-called {\em disorder potential\/} $\phi(\k)$
is added to the Hamiltonian
\be
\cH(\s|\k) \to \cH^\phi(\s,\k) = \cH(\s|\k) + \phi(\k)\ ,
\label{meth1}
\ee
which is then determined in such a way that configuration averaging,
as implied by Eq.\ (\ref{fqu}), becomes part of the Gibbs average 
in the enlarged phase space.

To utilize the method in practice, one still needs an explicit 
representation for $\phi(\k)$. For spin  or site diluted 
systems  the following representation in terms of products of occupation 
numbers 
\bea
-\beta\phi(\kappa) & = & \lambda_1 \sum_i (k_i -\rho) + \lambda_2 
\sum_{\langle i j \rangle} (k_i k_j - \rho^2) +\ \dots \nonumber\\
& & + \lambda_P\sum_P \left(\prod_{i\in P} k_i -\rho^{|P|}\right ) +\ \dots
\label{meth3}
\eea
is always suitable. The first sum explicitly displayed in (\ref{meth3}) 
is over all lattice sites, the second over all nearest neighbour pairs, the 
third over all elementary plaquettes (of size $|P|$) of the system, and so 
on.

It has been realized \cite{Ku96} that the self-consistent constraint 
equations which determine the couplings $\lambda_{\nu}$ as a function of 
temperature $T$, magnetic field $H$ and spin density $\rho$ (Eqs.\ 4 of Ref.\ 
\onlinecite{Ku94}), can be written as necessary conditions describing the 
{\em maximum\/} of the equilibrium ensemble's dimensionless free energy 
(per site) $f^\phi = - N^{-1} \ln Z^\phi = - N^{-1} \ln \sum_{\sigma,\k} 
\exp{ [ - \beta \cH^\phi(\s,\k)]}$ with respect to the $\lambda_{\nu}$'s.
They can therefore be written, and are in practice evaluated too, as
\bea
-\frac{\partial f^\phi}{\partial \l_1} &=& \ld k_i \rd_\phi-\rho = 0\ , 
                                                                 \nonumber \\
-\frac{\partial f^\phi}{\partial \l_2} &=& \ld k_i k_j \rd_\phi - \rho^2 
                                                    = 0\ ,\ \dots \nonumber \\ 
-\frac{\partial f^\phi}{\partial \l_P} &=& \Big\langle \prod_{i\in P} 
k_i \Big\rangle_\phi - \rho^{|P|} = 0\ ,\ \dots\ .                    
\label{meth4n}
\eea

Solving Eqs.\ (\ref{meth4n}) for every cluster of connected sites 
-- i.e. exactly obtaining the quenched free energy via the EEA -- is
clearly infeasible. However, a systematic scheme of approximations 
may be put up by matching only a {\em subset\/} of the full set 
of moments of 
the equilibrium ensemble's Gibbs distribution to the corresponding subset 
of moments of the problem with quenched disorder. 
This amounts to forcing $k_i$ 
correlations on larger and larger groups of lattice sites to coincide with 
those of the quenched system. 
The larger the number of constraints taken into account, the better a 
description of the fully quenched system is obtained.

In Ref.\ \onlinecite{Ku94} four different approximating systems 
named {\em (a) - (d)\/} were studied, 
differing by the number of terms kept in (\ref{meth3}), 
and thus by the set of constraints which is actually retained in 
(\ref{meth4n}). 
The present study is focused only on approximation {\em (d)}: 
the first three contributions to the disorder potential
are kept, i.e. single site, pair and square plaquette terms. 
In addition, due to the anisotropy of the strip geometry employed 
in the calculation (see Sec.\ \ref{tran} below), 
pair occupancy parallel and perpendicular to the
strip are treated as separate constraints, i.e. they are controlled 
by separate couplings  
$\l_{2\parallel}$ and $\l_{2\perp}$ in the disorder potential which,
in turn, have to be determined by two distinct equations of the 
type (\ref{meth4n}).

\subsection{Transfer Matrix Approach}
\label{tran}

The approach outlined above might 
-- at least in principle -- solve our problem to any
desired degree of accuracy. However, in $d\ge 2$ not even the simplest 
approximating system is exactly solvable. To analyse the thermodynamics or 
the critical behaviour of our 
systems, we use the strip FSS
techniques pioneered by Nightingale. \cite{Ni76} A distinct
advantage of the EEA in this context is that we 
are dealing with {\em translationally invariant equilibrium systems\/}
with short range couplings 
at all levels of 
approximation within the moment matching scheme 
described above. 
FSS can therefore be implemented using conventional TM techniques
for nonrandom systems.

Our strip is a $L\times L^{\prime}$ square lattices, with the
thermodynamic limit taken in the $L^{\prime}$ direction, 
and periodic boundary conditions (unless otherwise stated) imposed 
in the perpendicular direction.
The row to row transfer matrix $\Gamma$, constructed from 
the local Boltzmann factors, can be chosen to be symmetric for all 
systems {\em (a) - (d)\/} of Ref.\ \onlinecite{Ku94}, 
and is diagonalized for strips of fixed width $L$, with the appropriate
set of constraints (Eqs.\ (\ref{meth4n})).

The dimensionless free energy $f_L$ (per site) is related to the largest 
eigenvalue 
$\gamma_1$ of $\Gamma$ via
\be
- f_L =  L^{-1} \ln \gamma_1
\ee
(to simplify notation, we omit stating the $L$ dependence of $\G$ and its 
eigenvalues in what follows).

The constraint equations (\ref{meth4n}) 
have been formulated in 
terms of first derivatives of the free energy, and thereby in terms 
of first derivatives of $f_L$.
As mentioned 
above, those derivatives are nothing but gradient information in the 
problem of maximizing $f_L$ (or minimizing  $- f_L$) over the appropriate 
set of couplings. Notice that the transcendental nature of these equations
necessitates an iterative numerical solution. Having in this way
determined the (approximate) disorder potential, one may compute thermodynamic
functions and, in particular, the correlation length of the order parameter
fluctuations, which is needed in the phenomenological renormalization group
scheme. The correlation length is given in terms of the largest and the
second largest eigenvalue of the TM as
\be
\xi_L^{-1} = - \ln \frac{\gamma_2}{\gamma_1}\ .
\label{csi1}
\ee
Thermodynamic functions of interest are obtained by differentiating the 
free energy with respect to temperature or magnetic field. Here, care 
must be taken because of implicit field and temperature dependences in the 
couplings of the disorder potential. In other words, derivatives have to be 
performed {\em along the solution-manifold of the appropriate set of 
constraint equations.\/} 

Introducing modified spins $S_i = k_i \s_i$ with values in $\{0,\pm 1\}$,
in a single variable we 
encode, both, presence or absence of a spin
(the occupation numbers being given by $k_i = |S_i|$),
and the spin states. 
Since the dimension of $\G$ grows as $3^L\times 3^L$, 
we have used two different strategies -- with complementary 
strengths and weaknesses -- 
to simplify the computational task of dealing with huge transfer matrices.
The first is a group theoretical analysis which exploits the various
symmetries of the matrix and effectively reduces its dimensionality. 
The second is the use of sparse matrix techniques. 

We will not bother the reader by giving a detailed account of these techniques
since, apart from the fact that they have been non-trivially specialized
to deal with our particular system, they are not new (for the sparse
matrix technique see, e.g., Ref.\ \onlinecite{Ni79}).
We just notice that within either of them it is possible to identify the
blocks $\Gamma^{(1)}$ and $\Gamma^{(2)}$, in the block diagonal
representation of $\Gamma$, that contain its two largest eigenvalues
$\gamma_1$ and $\gamma_2$ respectively. $\Gamma^{(1)}$ is the block 
spanned by the space of eigenvectors transforming symmetrically under
spin reversal, while 
$\gamma_2$ is found in the block spanned by the eigenvectors of $\Gamma$
which are anti-symmetric under spin reversal.
In a similar spirit, one may enquire into
the large distance behaviour of the $k_i$ correlation functions $G_k(r_{ij})
=\langle k_i k_j\rangle - \langle k_i  \rangle \langle k_j \rangle$.
This requires to locate a corresponding eigenvalue $\tilde\g_2$, 
which controls the asymptotic decay of $G_k(r)$. 
It is found
as the second largest eigenvalue in the block corresponding to a representation
of the symmetry group which is 
{\em symmetric} with respect to both external 
(spatial) and internal (spin) 
symmetries of the system. Indeed, whereas the spins $\sigma_i$ change 
sign under global spin reversal, entailing that the eigenvalue to be inserted 
in (\ref{csi1}) for the computation of the spin-spin correlation length 
must belong to the antisymmetric block of $\Gamma$, the disorder variables 
$k_i = |S_i|$ do {\em not}. 

The computation of second order derivatives of $\gamma_1$,
needed to evaluate some of the relevant thermodynamical quantities, 
is feasible only
in the group theoretical reduction approach. The second derivatives in fact
require knowledge of the complete eigensystem, which is still far too
large within the sparse matrix approach. In this case one has to
resort to finite differences of first order derivatives to compute,
e.g., the specific heat or the spin susceptibility. 
The advantage of the sparse-matrix approach, on the other hand, is that 
it can be
pushed to considerably larger strip-widths than the group theory approach. 
We reach the value $L=9$ with the latter and $L=13$ with the former.
Moreover, this approach does not exploit any symmetries 
beyond the spin reversal symmetry in zero magnetic field. 
It is therefore also applicable
if one introduces an anti-ferromagnetic seam along the strip, 
a device which can be used to compute the domain wall free energy. We 
shall exploit this quantity below to considerably boost the precision of 
our FSS analysis.


\section{The Two Scenarios}
\label{scen}

Before showing our results,
it is time to briefly present the two main scenarios which have survived 
over a wider range of possibilities for the description of the 
critical behaviour of the 2$d$ site diluted Ising model,
but which are 
obviously mutually excluding.

The {\em logarithmic corrections\/} scenario is based on the
quantum field theory results by Dotsenko and Dotsenko \cite{DD83}
and Shalaev, \cite{Sha84} Shankar \cite{Shan87} and Ludwig \cite{Lu88}
(the latter contributions correcting certain errors in the former).
Though strictly valid only in the limit of weak disorder,
they indicate that the presence of impurities affects the
critical properties of the model only through a set of 
logarithmic corrections to the pure system behaviour.
In particular, according to this picture, the correlation
length of the infinite system close to the phase transition 
is expected to show the form 
\be
\xi_{\infty} \sim t^{-\nu} \left( 1 + \tilde g \ln \left( 
                            \frac{1}{t} \right) \right) ^{\tilde{\nu}} \, ,
\label{csilog}
\ee
where $t$ is the reduced temperature $t=(T-T_c)/T_c \ll 1$ 
($T_c$ being the critical temperature, which will depend on $\rho$),
the exponents $\nu$ and $\tilde{\nu}$ are respectively $1$ and 
$1/2$, and $\tilde g$ is a non-negative constant, such that 
$\tilde g=0$ in the pure case (usual power law behaviour recovered) 
and increases with increasing disorder. 
A similar behaviour holds for the magnetic susceptibility:
\be
\chi_{\infty} \sim t^{-\gamma} \left( 1 + \tilde g \ln \left( 
                        \frac{1}{t} \right) \right) ^{\tilde{\gamma}} \, ,
\label{chilog}
\ee
with $\gamma=7/4$ and $\tilde{\gamma}=7/8$.
For the specific heat, too, the exponent of the pure case 
$\alpha=0$ is not modified by the introduction 
of disorder, but the simple logarithmic behaviour 
$C_{\infty} \sim  \ln (1/t)$ is substituted by the double
logarithmic singularity
\be
C_{\infty} \sim \ln \left( 1 + \tilde g \ln \left( 
                        \frac{1}{t} \right) \right)  \, .
\label{cvlog}
\ee
The behaviour of the critical spin-spin correlation function 
is instead predicted not to change in presence of impurities,
thus its anomalous dimension $\eta$ retains its pure Ising value
$\eta=1/4$.

Recent numerical works \cite{Fae92,KiPa94,Ku94}
have provided evidence in contrast to what 
has just been presented above.
These findings show quantities such as the 
susceptibility and the correlation length to display 
simple power-law singularities, at criticality, with exponents 
$\gamma$ and $\nu$ varying continuously with disorder
in a way, however, that their ratio $\gamma/\nu$ is
kept constant at the pure system value.
According to this {\em weak universality\/} \cite{Suz74} scenario,
the ratio of exponents should not depend on disorder, and 
the specific heat was observed to saturate at a nondivergent
value as $t \to 0$.

In a finite-size numerical investigation
like ours it is also essential to translate the predictions of 
these scenarios into their finite-size counterparts. This is
straightforward and well known in the case of power law singularities 
(and power law corrections to scaling), but less trivial within
the logarithmic corrections scenario. These questions are dealt 
with in detail in the Appendix and, where they arise, in the 
following Section.

As already stressed in the Introduction, 
we have tried to look at our results with open mindedness:
none of the two scenarios is chosen as a reference, 
we rather attempt to let our analysis to decide for those
of the two which consistently fits
the whole series of our data.

\section{Main results}
\label{resu}

\subsection{Determination of the critical temperature}

The 2$d$ spin diluted Ising model 
exhibits a phase transition from a paramagnetic to a ferromagnetic 
low-temperature phase, provided the spin density $\r$ is 
larger than the critical site percolation density  
$\rho_c \simeq 0.592745$. \cite{perc}
The model unfortunately lacks the 
possibility, known for its random-bond counterpart,
to exactly determine the critical temperature 
as a function of the degree of disorder
from duality.
This is one of the reasons which render the bond disordered
version 
more attractive for a numerical study, i.e. the possibility of
studying its critical behaviour by sitting exactly {\em at\/} $T_c$.
A very precise numerical 
determination of $T_c$ is thus called for in our
case, and it has indeed been reached by a joint analysis of the
finite-size behaviour of the correlation length (as defined
in Eq.\ (\ref{csi1})) and of the {\em domain wall free energy\/}
(or {\em interfacial tension\/}), the latter obtained by 
comparing free energies of systems with periodic and
antiperiodic boundary conditions in the direction 
perpendicular to the strip.
Antiperiodic boundary conditions in a ferromagnetic system force  
an interface along the cylinder's length.
In practice, the interface is created by 
introducing an antiferromagnetic seam along the cylinder, i.e. 
by reversing in each row the bond $ J \to - J$ between two fixed spins
(e.g. spins $L$ and $1$).
The interface (domain wall) free energy per unit length $\sigma_L$
is given by the difference in free energy between the system with
periodic and antiperiodic boundary conditions, and reads
\be
\beta \sigma_L = - \ln \frac{\gamma_{1, abc}}{\gamma_1}\,
\label{wall1}
\ee
with $\gamma_{1,abc}$ the largest eigenvalue of the TM with
antiperiodic boundary conditions (compare Eq.\ (\ref{wall1}) to 
Eq.\ (\ref{csi1})).

\begin{figure}
\vspace{-0.4truecm}
{\centering 
\epsfig{file=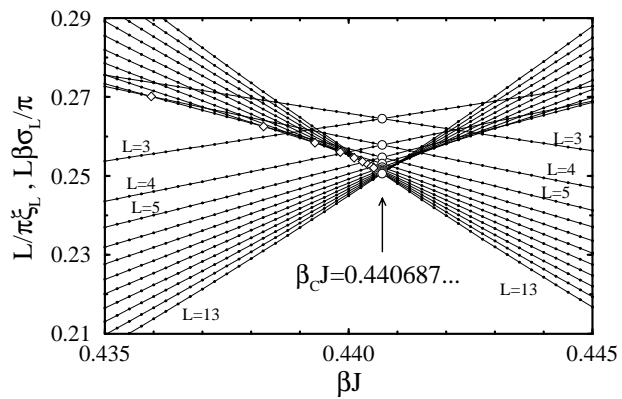,width=0.400\textwidth}  
\par}
\vspace{0.2truecm}
\caption[]{
Set of curves $L/\pi\xi_L$ (negative slope) and
$L\beta\sigma_L/\pi$ (positive slope) as a function of
$\beta J$ for different values of $L$, for the pure Ising model.
Their mutual intersections, at the absciss\ae \ 
$(\beta J)_L^{int}$, are indicated by open circles, the crossings
of two $L/\pi\xi_L$ curves for couples of values
$L$, $L+1$ ($(\beta J)_L^{\xi}$) are drawn as open diamonds,
while black dots denote the points used in the temperature scan.
Is is evident how the former sequence is much more rapidly converging
than the latter to the exact value 
$\beta_c J = \ln(\sqrt{2} +1)/2 = 0.440687 \dots\,\,$. The value of 
the limit of both sequences of points on the vertical axis is the
exponent $\eta = {1}/{4}$.
}
\label{cros}
\end{figure}

Phenomenological renormalization \cite{Ni76,Ni79} predicts that the
correlation length $\xi_L$ scales as $L$ at criticality. 
The correct formula reads 
\be
\xi_L^{-1}  =  L^{-1} \left( A + B_{\xi}    L^{y_{irr}} + \dots \right)
\,\, , 
\label{csi2}
\ee
where the correction-to-scaling terms (the exponent $y_{irr}$ 
is related to the presence of irrelevant scaling fields and 
it is known to be $y_{irr} = -2$ for the Ising 
model \cite{Blo-den}) split the exact crossing of the curves 
${L}/{\xi_L}$ versus $\beta J$ at the critical point
for different values of $L$ into a sequence of distinct 
intersections of absciss\ae\ $(\beta J)_L^{\xi}$
(see Fig.\ \ref{cros}).
Extrapolating this sequence 
to its asymptotic value provides an accurate determination of 
the critical temperature.
The task is accomplished 
by a combination of different extrapolation techniques.
They are basically: the Bulirsch and Stoer
algorithm discussed by Henkel and Sch\"utz; \cite{Henkel} 
the three-point iterated fit method presented by Bl\"ote 
and Nienhuis; \cite{Blo-Ni} 
a third method, close in spirit to the latter, which 
uses fitting procedures at all stages of 
extrapolation. \cite{citeitornot} 
These three different algorithms
give accurate and consistent results, since they
lead to extrapolated values which do not significantly
differ from each other. 
They are also used in the following wherever 
the  $L \to \infty$ limit of a sequence needs to be extracted.
Their comparison sets the error bars of our analysis.

The same scaling argument presented for $\xi_L^{-1}$ can be
applied also to the wall free energy 
$\beta \sigma_L \sim L^{-1}$, 
and the corresponding sequence 
$(\beta J)_L^{\sigma}$ 
similarly analyzed. 

\begin{figure}
{\centering 
\epsfig{file=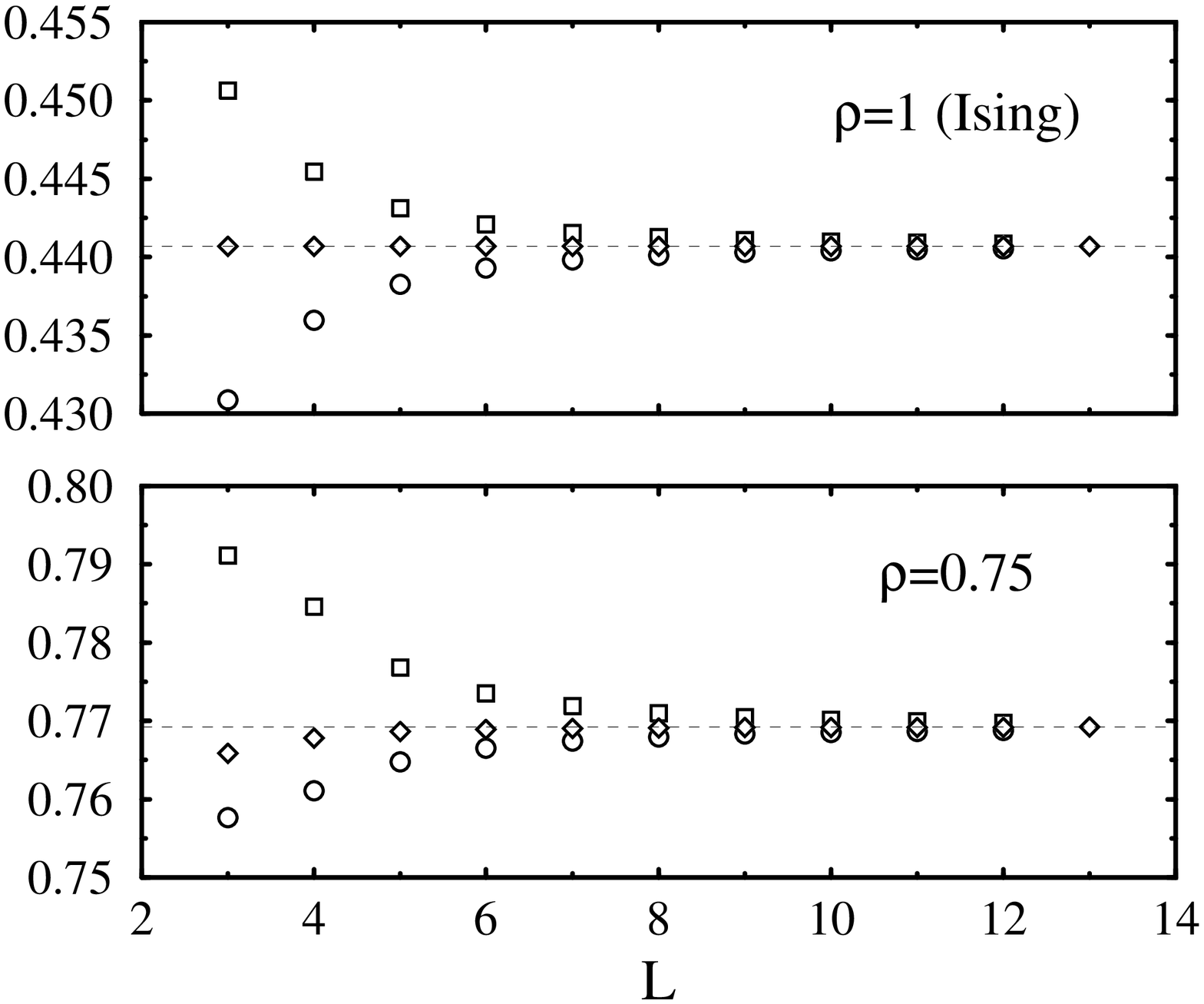,width=0.400\textwidth}  
\par}
\vspace{0.2truecm}
\caption[]{
The sequences of finite-size approximations to the critical
inverse temperature 
$(\beta J)_L^{\xi}$ (circles),
$(\beta J)_L^{\sigma}$ (squares) and 
$(\beta J)_L^{int}$ (diamonds)
as a function of strip-width $L$ 
for $\rho=1$ (the pure Ising case) and for spin density $\rho=0.75$.
It is apparent that in both cases the last sequence converges faster 
than the other two to the critical inverse temperature
of the infinite system represented by the dashed line.
Here, as well as in the following pictures, no error bars are
shown if smaller than the symbol's size.
}
\label{conv}
\end{figure}

The apparent ``mirror symmetry" of the two sets of curves
displayed in Fig.\ \ref{cros} suggests also a new method to extract
$\beta_c J$: the analysis of the sequence 
$(\beta J)_L^{int}$ of mutual intersection points of the
curves $L/\xi_L$ and $L \beta \sigma_L$ versus $\beta J$ for
fixed values of $L$.
This sequence also converges to $\beta_c J$ in the limit of large 
$L$, but {\em much\/} faster than the previous sequences
$(\beta J)_L^{\xi}$ and $(\beta J)_L^{\sigma}$,
and its extrapolation 
through the methods described above furnishes a much more precise 
determination of the critical temperature (upper half of Fig.\ \ref{conv}).
Already for the rather small value $L=6$, 
the pure 2$d$ Ising model value of $(\beta J)_L^{int}$
differs from the exact critical value only after the 11th decimal digit. 
Neither this striking result nor the analysis of the
$(\beta J)_L^{int}$ sequence are, to our knowledge, present 
in the literature, and we regard them as remarkable 
findings on their own.
 
The whole machinery can be carried over to the diluted case. 
It is worth remembering that in our investigation 
the spin density $\rho$ is treated as a parameter 
fixed to definite values in the range 
$\rho_c \leq \rho \leq 1$
to scan the phase diagram.
The values of $\rho$ analyzed are 
$\rho = 0.95, \, 8/9, \, 0.8, \, 0.75, \, 2/3 $. 

For $\rho \neq 1$ 
the ``mirror symmetry"
of Fig.\ \ref{cros} is lost when the value of
$\rho$ substantially departs from the pure system value. 
Logarithmic terms would also appear in Eqs.\ (\ref{csi2}) 
in case of the logarithmic corrections scenario 
\cite{Car86} (see the Appendix and the discussion in 
Sec.\ \ref{etagamma} below).
In spite of that, the sequences $(\beta J)_L^{int}$ 
still converge quite fast to their asymptotic value, 
again allowing extremely accurate determination of 
$(\beta_c J )(\rho)$ 
(see lower half of Fig.\ \ref{conv})
also for 
the lower values of $\rho$, where closeness to the 
percolation threshold induces stronger finite-size
deviations from the asymptotic behaviour of the 
quantities under investigation.

\begin{figure}
\vspace{-0.4truecm}
{\centering 
\epsfig{file=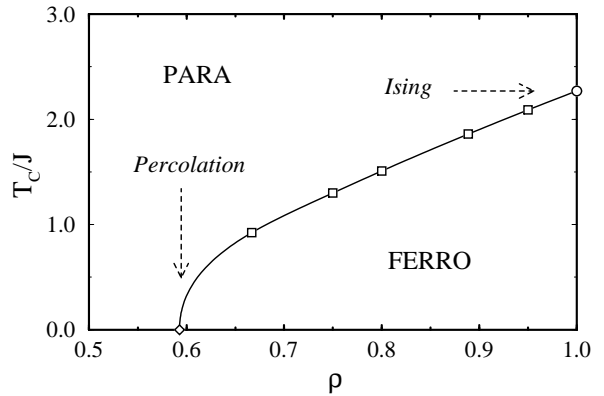,width=0.400\textwidth}  
\par}
\vspace{0.2truecm}
\caption[]{
The phase diagram of the 2$d$ site diluted Ising model,
with its ferro- and paramagnetic phases.
The diamond and the circle denote respectively 
the exact percolation threshold and the transition
temperature of the pure Ising model, the squares correspond 
to the values of spin density $\rho$
analyzed in this investigation.
The line is nothing but a guide to the eye.
}
\label{phased}
\end{figure}

Fig.\ \ref{phased} presents the phase diagram of the
2$d$ site diluted Ising model, showing the values of 
$T_c(\rho)/J$ obtained in this way
for strip-widths up to $L = 13$.

The critical slope of the curve $T_c(\rho)$ versus $\rho$ 
at $\rho=1$ is exactly known \cite{Tho} to be 
$S_c = T_c^{-1} dT_c/d\rho|_{\rho = 1} = 2/[\ln (1+\sqrt{2}) (1+\sqrt{2}/\pi)]
                                       = 1.564785\dots\ $;
from an additional calculation limited to size $L=11$
at a value of $\rho$ very close to $1$ and
from a numerical evaluation of the derivative, 
we obtain an estimate which differs
from the exact result by approximately $0.01\%$. 

Let us finally point out that 
the method just described provides a very precise determination
of the critical temperatures 
{\em for each of our approximating systems.\/} 
We make no claim, however, that these values coincide with
those of the fully quenched system: for fixed $\rho$
they in fact turn out to be slightly different from
system {\em (a)\/} to {\em (d)\/} \cite{Ku94}
and need not be the same as Monte Carlo data. \cite{Ball+97}
The reason is that nonuniversal quantities, such as indeed 
the critical temperature or the value of the percolation
threshold $\rho_c$, do depend on the approximating system
chosen, while universal quantities, such as critical exponents,
central charge, etc\dots, do not. Our confidence in this statement,
and consequently in the validity of our method, will be addressed 
in Sec.\ \ref{disc}. 

\subsection{The exponent {\mbox{\boldmath ${\eta}$}} 
            and the ratio {\mbox{\boldmath ${\gamma/\nu}$}} }
\label{etagamma}

Having determined with great precision the critical temperature of our model,
we now turn to the evaluation of critical exponents.

The theory of conformal invariance \cite{Cardy}
relates the amplitude $A$ of the leading term in the FSS
behaviour of $\xi$ (Eq.\ (\ref{csi2})) to the anomalous 
dimension of the spin-spin correlation function via
$ A = \pi \eta $,
providing 
a simple method to extract $\eta$ just by extrapolating
the sequence $L/\xi_L$ to $L \to \infty$.
The same procedure can be applied as well to $L \beta \sigma_L$ since 
the amplitudes for the correlation 
length and for the wall free energy are equal.

In the pure Ising case, $\eta$ is known to be ${1}/{4}$.
Our extrapolations unambiguously give the same value 
\be
\eta = \frac{1}{4}
\label{eta}
\ee
for all the values of $\rho$ considered, with a maximum estimated error
of $0.0004$ (the error arising from the uncertainty in the 
location of the critical temperature being much smaller than
the difference between final values from different 
extrapolation procedures). From this we can conclude that the exponent 
$\eta$ does not depend on disorder, a result which has been known already
\cite{KiPa94,Ku94,Sel+94,sldq95,sldq97} but without the precision shown 
above.

\begin{figure}
\vspace{-0.4truecm}
{\centering 
\epsfig{file=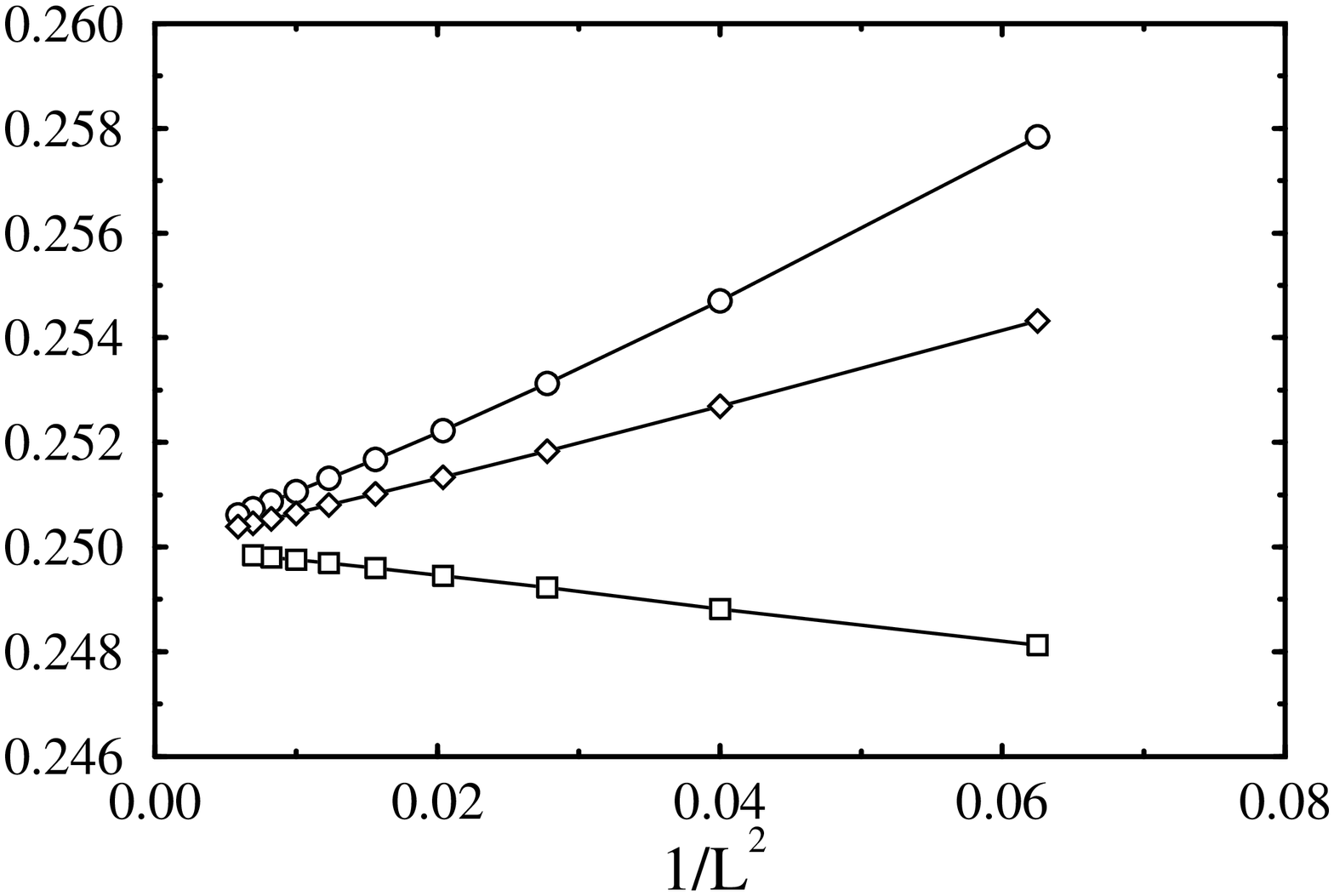,width=0.400\textwidth}  
\par}
\vspace{0.2truecm}
\caption[]{
The scaled inverse correlation length $L/ \pi \xi_L$ plotted as a 
function of $1/L^2$,
for the pure Ising case (circles), for $\rho=0.80$ (diamonds)
and for $\rho=2/3$ (squares).
The lines just connect the displayed points.
Note the vertical scale.
}
\label{ellesucsi}
\end{figure}

Note that the correlation length $\xi$ extracted from the 
ratio of TM eigenvalues (Eq.\ (\ref{csi1}))
within the EEA refers to the 
decay of the {\em average\/} spin-spin correlation function 
in the disordered system. 
This is to be contrasted with the correlation length 
that describes the {\em typical\/} decay of the correlation 
function in a given disorder configuration, a quantity readily 
obtainable from a corresponding ratio
of Lyapunov exponents within a random TM approach. \cite{sldq95,sldq97}      
Our identification may be confirmed by plotting $L/\xi_L$ versus $1/L^2$ 
(Fig.\ \ref{ellesucsi})
and noticing that no corrections to a linear dependence
smoothly extrapolating to ${1}/{4}$ appear, as it 
could instead
be expected for the behaviour of the typical correlation length
(see in particular the discussion focusing on Fig.\ 1 of 
Ref.\ \onlinecite{sldq95}, and Ref.\ \onlinecite{sldq97}).

We would like to point out that the smallness of the
estimated error of the exponent $\eta$ 
is of particular relevance. It can in fact
be also read as a strong confirmation that our
approximations are sufficient to place the model under study 
in the right universality class, and that universal 
quantities therefore turn out to be correctly
reproduced.

This is confirmed by the behaviour of the magnetic 
susceptibility data obtained via the group theoretical 
reduction technique. The finite size scaling analysis of the 
susceptibility $\chi$ predicts a leading $L$ dependence of 
the form $\chi_L \sim L^{\gamma/\nu}$, and the ratio $\gamma/\nu$
is extracted by analyzing the sequence
\be
\left( \frac{\gamma}{\nu} \right) _L = 
                         \frac{\ln \left( \chi_{L+1}/\chi_L \right)}
                              {\ln \left( L+1/L       \right)}\ .
\label{estimates}
\ee
Corrections
to scaling generally modify the pure power law, and an extrapolation 
to $L\to \infty$ is needed. Corrections may be again of pure power 
law form 
in $L$ (in the conventional situation arising from irrelevant scaling 
fields), or indeed by terms containing logarithms of $L$ (in the 
log-corrections scenario). In the latter case, if the largest 
available 
system size is too small, the data might still reflect a 
preasymptotic regime, and the very idea of extrapolating a sequence
like (\ref{estimates}) may lose its validity in that case.
We refer the reader to the Appendix for a discussion of this point.
We are however not able to
detect the non-universal corrections predicted to be present in the
preasymptotic regime of the log-corrections scenario. 
They are either absent, or too small to be detected
(Eq.\ \ref{pregam}).
The same is true for $\eta$.

Extrapolating the sequence (\ref{estimates}) provides the result 
\be
\frac{\gamma}{\nu} = \frac{7}{4}
\label{gammaovernu}
\ee
for all the points studied in the present paper.
The estimated error is around $0.001$, bigger than for $\eta$ because 
a shorter sequence ($L_{\rm max}$ = 9) 
has been utilized. These data show that the ratio $\gamma/\nu$
is independent of the degree of dilution, a finding in complete 
accordance with the constancy of $\eta$ via the Fisher's relation 
$\gamma/\nu = 2 - \eta$, which is thereby demonstrated to be satisfied 
with a precision of the order of a tenth of a per cent. 

The observed constancy of $\eta$ and $\gamma/\nu$ is unfortunately not 
sufficient to discriminate between the two admissible scenarios, since 
{\em both\/} predict that the ratio retains the pure Ising value $7/4$, 
irrespective of disorder. It only shows that for these two quantities 
extrapolations are not noticeably misguided by logarithmic corrections, 
in case they are present.

\subsection{The central charge} 

Conformal invariance \cite{Blo-Car} provides a link between 
the values of the dimensionless free energy per site 
at criticality $f_L$ in a finite strip with periodic boundary 
conditions, and the central charge $c$ which characterizes the
universality class of a conformally invariant model. 
The relation reads
\be
f_L \sim f_{\infty} - \frac{\pi c}{6 L^2} +\dots \, ,
\label{cc}
\ee
with $f_{\infty}$ the nonuniversal bulk (infinite system) 
free energy per site. Eq.\ (\ref{cc}) is supposed to retain its
validity also in a random system, provided $c$ is substituted 
by an {\em effective\/} central charge $c^{\prime}$ 
(Ref.\ \onlinecite{Lud-Car}). The next-to-leading term, omitted in
(\ref{cc}), differs 
depending on whether logarithmic corrections are present or not.
In the latter case it is believed to depend on $L$ as
$B_f L^{-2+y_{irr}} = B_f L^{-4} $ if we adopt the value 
$y_{irr} = -2$ of Ref.\ \onlinecite{Blo-Ni} or, equivalently, 
that born out from the analysis of Sec.\ \ref{etagamma} above.
In the former case, instead, Ludwig and Cardy \cite{Lud-Car} 
derived the results presented in Eqs.\ (\ref{cL}) of the Appendix.
To avoid dealing with the $f_{\infty}$ term, we prefer to extract
finite-size estimates of $c^{\prime}$ through the formula
\be
c^{\prime}_L = \frac{6}{\pi} \left(f_{L+1} - f_L \right)
               \frac{L^2 (L+1)^2}{2L + 1}
\label{ccl}
\ee
and analyze them both by extrapolating the above sequence to
$c^{\prime}$, and possibly looking for finite-size 
corrections (next-to-leading terms).

The outcome of our analysis does not sensibly differ from what
we obtained in the study of the exponent $\eta$. That is,
while $c^{\prime}$ appears to be clearly given by
\be
c^{\prime} = \frac{1}{2} \, , 
\label{cprime}
\ee
equal to the pure Ising value, for all the points studied 
(with a maximum error of $0.0003$), we are not 
able to individuate any correction of the kind of those 
expressed by Eq.\ (\ref{cL}). The plots of $c^{\prime}_L$ 
versus $1/L^2$ are lines showing no appreciable
bending.  

The estimates $c^{\prime}_L$ converge to their asymptotic value 
always from above, both in the pure as well as in the 
diluted case. This is at least a necessary condition for 
the reflection positivity property,  
which holds for unitary models but it might not hold
for random models. The discussion on this point in Ref.\ 
\onlinecite{Jac-Car} is indeed based on Eqs.\ (\ref{cL}) which 
would predict a convergence from below, the first correction terms 
to the value $c'=1/2$ being negative. But again, they are either 
absent, or too small to be discernible within our errors, or
convergence from below sets in for system sizes beyond those
accessible by our TM calculations.

\subsection{The analysis of the correlation length}

The correlation length provides us with 
another tool to push our investigation further:
the study of the exponent $\nu$,
extracted, in addition to $\eta$, from a 
FSS analysis of $\xi$, can be of particular
relevance since it is a single isolated 
exponent, therefore it is predicted to show
a different behaviour in the two scenarios.

Our analysis follows general reasoning. \cite{Ni76}
A recent application to disordered Ising models
has been provided by Aar\~{a}o Reis {\em et al.\/}.
\cite{Aa+97} 
Two quantities at our disposal can be subjected to the 
same investigation, the correlation length $\xi$
as well as the domain free energy $\beta \sigma$: 
both show the same asymptotic behaviour, though with
different corrections to scaling.

First we define the derivatives of these
quantities with respect to temperature:
\begin{eqnarray}
\mu(t)          & = & \frac{d \xi(t) }{dt}    \, ,  \\
\mu^{\prime}(t) & = & \frac{d \left[ \beta \sigma(t) \right] }{dt}^{-1} \, .   
\label{muoft}
\end{eqnarray}
A pure power-law divergence of $\xi(t)$ at criticality like 
$t^{-\nu}$ implies $\mu(t) \sim t ^{-\nu- 1}$, which in turn 
translates into the finite-size prediction $\mu_L \sim L^{1+1/\nu}$. 
The same relations hold respectively for $\beta \sigma(t)$, 
$\mu^{\prime}(t)$ and $\mu^{\prime}_L$. From the $\mu_L$ sequence 
(and similarly form the $\mu'_L$ sequence), one obtains 
finite-size approximations $\nu_L$ of the correlation 
length exponent via
\be
\nu_L^{-1} = \frac{\ln \left( \mu_{L+1}/\mu_L \right)}
                  {\ln \left( L+1/L       \right)} - 1 \ .
\label{nuofl}
\ee
Extrapolated values 
clearly appear to vary with $\rho$, and to change continuously from 
the Ising value $\nu = 1$ to the percolation value 
$\nu = 4/3$, \cite{denNijs} in a way shown in Table\ \ref{expo}. 
The large error bars on the values corresponding to stronger dilution 
are due to degradation in the quality of the fits, and even to 
the appearance of non-monotonicities in the sequence $\nu_L$ versus 
$L$ extracted from the domain wall free energy data,
which prevent an extrapolation as precise as those for
larger $\rho$ values. The cause is probably the closeness of the 
percolation threshold, which requires a corresponding enlargement
in system size in order to disentangle the 
system's correct critical behaviour from crossover effects.

\begin{table}[t!] 
\caption{
Various exponents for different values of spin density $\rho$. 
In the second column the exponent $\nu$ as extrapolated 
from Eq.\ (\ref{nuofl}). The third column shows the value
of the ratio $\alpha/\nu$ extracted from Eq.\ (\ref{cvofl2}). 
The last column provides a test for the validity of the 
hyperscaling relation $2/\nu- \alpha/\nu=d=2$.}
\begin{tabular}{cccc} 
$\rho$ & $\nu$ 
& 
$\phantom{\frac{\alpha}{\nu_{8_8}}}\,\,$ 
$\frac{\alpha}{\nu}$ 
$\phantom{\frac{\alpha}{\nu_{8_8}}}$ 
& 
                                    $\frac{2}{\nu} -\frac{\alpha}{\nu}$ \\
\tableline
0.95   & 1.054 $\pm$ 0.002 & -0.102 $\pm$ 0.008 & 1.999 $\pm$ 0.012     \\
8/9    & 1.113 $\pm$ 0.001 & -0.214 $\pm$ 0.007 & 2.011 $\pm$ 0.009     \\
0.80   & 1.15  $\pm$ 0.03  & -0.30  $\pm$ 0.02  & 2.04  $\pm$ 0.06      \\
0.75   & 1.18  $\pm$ 0.02  & -0.35  $\pm$ 0.02  & 2.05  $\pm$ 0.05      \\
2/3    & 1.23  $\pm$ 0.05  & -0.445 $\pm$ 0.005 & 2.07  $\pm$ 0.07	\\ 
\end{tabular}
\label{expo}
\end{table} 

A similar data trend was extracted also from Monte Carlo simulations, 
\cite{KiPa94} and analogous variation with the strength of 
bond disorder was observed in Ref.\ \onlinecite{Aa+97}.
The variation of $\nu$ as a function of $\rho$ 
at constant $\eta$ and constant $\gamma/\nu$ would qualify 
the observed non-universality --- if it must be upheld ---
to be of the {\em weak} form.

On the other hand, one should be wary of the fact that preasymptotic 
logarithmic corrections might invalidate the very idea of extrapolating 
a $\nu_L$ sequence of limited length; see (Eq.\ \ref{prenul}). This was 
the attitude advocated by Aar\~{a}o Reis {\em et al.\/}, who did not 
give much weight to an analogous variation of {\em their\/} extrapolated 
exponent $\nu$ as a function of disorder. Rather they proposed to fit 
the $L$ dependent $\mu_L$ data to a form
expected to hold in an intermediate (preasymptotic) range of $L$ values
within the logarithmic corrections scenario, viz. 
\be 
\mu_L \sim L^2 ( 1 - A \ln L )^{\frac{1}{2}} \, ,
\label{muofl}
\ee
which predicts $(\mu_L/L^2)^2$ to be linear in $\ln L$. 
Note that this expression used by Aar\~{a}o Reis {\em et al.\/}
was derived from heuristic considerations. It formally 
agrees with the true expression (\ref{premul}) only 
to first order in $\ln L$, but it predicts the {\em wrong\/} sign for the coefficient of the leading $\ln L$ 
behaviour of the effective size-dependent correlation length exponent $\nu_L$ 
(Eq.\ \ref{prenul}).

\begin{figure}
\vspace{-0.4truecm}
{\centering 
\epsfig{file=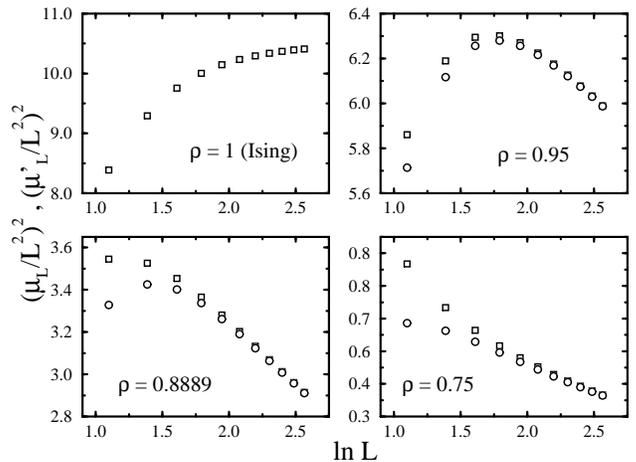,width=0.400\textwidth}  
\par}
\vspace{0.4truecm}
\caption[]{
The quantities $(\mu_L/L^2)^2$ and $(\mu^{\prime}_L/L^2)^2$
at criticality 
plotted as a function of $\ln L$, for various values of $\rho$.
}
\label{dcsi} 
\end{figure}
In Fig.\ \ref{dcsi} we plot our results in the same form as Aar\~{a}o Reis 
{\em et al.\/} do, both for $\mu_L$ and for $\mu_L^{\prime}$. The data 
show exactly the same trend as observed in Ref.\ \onlinecite{Aa+97}, both 
with respect to system size $L$ and with respect to the strength of the 
disorder. From this observation we derive additional strong confidence in 
the validity of our method. The pure system behaviour is soon substituted by 
an apparently linear term in $\ln L$ for $(\mu_L/L^2)^2$ in the 
increasingly disordered system, as would be expected from (\ref{premul}). 
Notice, once 
more, the same asymptotic behaviour of $\mu_L$ and $\mu_L^{\prime}$, 
with different corrections to 
scaling.

The data can, however, be consistently interpreted within a scenario 
of nonuniversal critical exponents as well. Indeed, for a power law 
divergence of the correlation length with $\nu > 1$, FSS would 
predict $(\mu_L/L^2)^2 \sim L^{-\omega}$ with $\omega = 2 - 2/\nu$ 
at sufficiently large $L$ for the data plotted in Fig.\ \ref{dcsi}. 
In the disorder range considered, $\omega$ is a rather small quantity 
as the $\nu$ extrapolated from Eq.\ (\ref{nuofl}) only slightly exceed 1. 
The data might therefore easily look like exhibiting a $(1 - A \ln L)$ 
behaviour for the accessible range of system sizes, which was the form 
assumed in Ref.\ \onlinecite{Aa+97}. This might however just be the result 
of an expansion to first order in the small quantity $\omega \ln L$ 
($L^{-\omega} \sim 1 - \omega \, \ln L$). 
Upon closer inspection 
a curvature compatible with an $L^{-\omega}$ behaviour is indeed discernible 
in the figure for larger disorder. Using a fit against $L^{-\omega}$ we 
have another way of analyzing our data to determine $\nu$, and the values 
obtained this way are compatible  with those obtained from extrapolation 
of the $\nu_L$ data. Nonetheless,
a curvature of the $(\mu_L/L^2)^2$ data may occur also 
in the log-corrections scenario 
(depending on the value of $D_1$ in Eq.\ (\ref{muL})). 

We can however affirm with certainty that neither Aar\~ao Reis 
{\em et al.} nor we are seeing preasymptotic effects of logarithmic 
corrections in the data for $\nu_L$, 
because they 
are {\em increasing\/} with $L$ (in our 
case at least 
for $\rho  > 0.8$), whereas according to FSS they should {\em decrease\/}; 
see 
Eq.\ (\ref{prenul}). Thus, {\em if\/} logarithmic corrections are present, 
they 
are for the available system sizes still masked by other corrections to 
scaling. Again, it seems that further investigation is required in order 
to be able to take sides on the critical behaviour of this system.

\subsection{The specific heat} 

The logarithmic correction scenario predicts for the specific 
heat the double logarithmic divergence of Eq.\ (\ref{cvlog}).
This translates, at criticality, into the FSS prediction \cite{Sel+94}
\be
C_L \sim C_0^{\prime} + C_1^{\prime} \ln \left(1 + C_2^{\prime} \ln L \right)
\, , 
\label{cvofl1}
\ee
where the pure system critical behaviour is recovered by the vanishing
of $C_2^{\prime}$ (but with $C_1^{\prime} C_2^{\prime} = const.$)   
for $\rho=1$.
On the other hand, a scenario of nonuniversal critical exponents
would lead to
\be
C_L \sim C_{\infty} + C_3^{\prime} L^{\alpha/\nu} \, .
\label{cvofl2}
\ee
Though $\alpha/\nu$ is a ratio of exponents, it is expected to vary with 
$\rho$, being directly related to $\nu$ via the hyperscaling
relation $2 -\alpha = d \nu$, with $d$ the system's dimensionality.
Moreover, this relation imposes a severe constraint on $\alpha$:
since $\nu$ is found to be greater than $1$, $\alpha$ should
be smaller than $0$, in other words the specific heat 
should then turn out to be nondivergent, with its finite-size estimates
saturating to a value $C_{\infty}$ at $T_c$. 

\begin{figure}
\vspace{-0.4truecm}
{\centering 
\epsfig{file=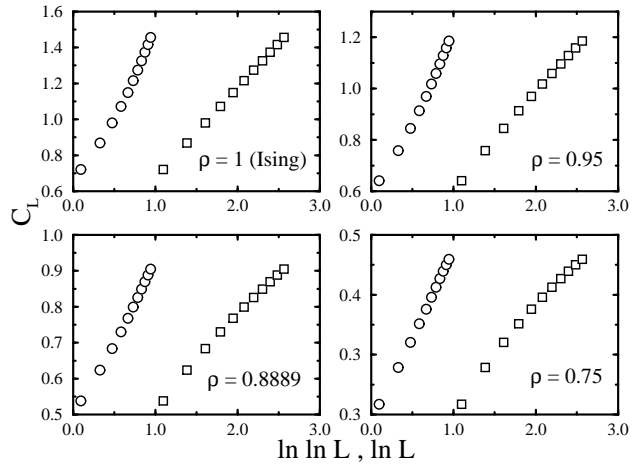,width=0.400\textwidth}  
\par}
\vspace{0.4truecm}
\caption[]{
The finite-size approximants of the specific heat
at criticality, $C_L$, versus $\ln \ln L$ (circles) and $\ln L$ (squares).
}
\label{cv} 
\end{figure}

We plot our data for the specific heat in the same form as in Ref.\
\onlinecite{St+97}, i.e. against both $\ln L$ and $\ln \ln L$ (Fig.\ 
\ref{cv}). The pure system divergence is well reproduced by the straight 
line the data show when plotted against $\ln L$, equivalently by the 
upward curvature 
when against $\ln \ln L$. Exactly as it happens in Ref.\ 
\onlinecite{St+97}, as disorder is switched on and increased, it is 
apparent that 
both curves tend to bend downwards, the former markedly deviating from,
the latter instead approaching, a straight line 
(from top left to bottom right of Fig.\ \ref{cv}).
The law describing these finite size data has therefore to
be at least less divergent with $L$ than $\ln L$.
However, the same 
kind of behaviour must at moderate system sizes be expected in {\em both\/}
scenarios (as we shall argue in greater detail in the concluding section).
It is therefore difficult to conclude definitely in favour of
Eq.\ (\ref{cvofl1}): here more than elsewhere the need for 
significantly larger system sizes is particularly felt.

If we fit the data to a nondiverging size dependence, in 
accordance with Eq.\ (\ref{cvofl2}), there is a consistency 
check to carry out, i.e. one may ask oneself whether the 
hyperscaling relation is verified or not, given the 
$\nu$ values from the correlation length.
The fitting of the specific heat to the form (\ref{cvofl2})
is performed first by taking all the points into account, 
then by reducing the data set by successively increasing the
initial value of $L$ considered. In this way one can both check
for the stability of such a fit and get a sequence of values
of $\alpha/\nu$ which may be subjected to subsequent 
extrapolation procedures, when needed. The results of this
analysis are shown in Table\ \ref{expo}. Its main outcome 
is the fact that, within a weak universality
interpretation of the data, a small, seemingly systematic offset
from hyperscaling, though still within estimated error 
bars, appears.

We have tried to discriminate between Eqs.\ (\ref{cvofl1}) and 
(\ref{cvofl2}) on the basis of a $\chi^2$ test. Although such 
a test appears to favour the log-corrections scenario by roughly 
an order of magnitude in $\chi^2$, the $\chi^2$ values for both
alternatives are so small 
(typically $\cO(10^{-8})$ and $\cO(10^{-7})$ respectively
for a choice of the data set) 
as to make it doubtful whether a 
meaningful model selection should be based on them. 
In particular, our data are not 
normally distributed random data
(indeed they are deterministic), 
as instead strictly required by a reliable
$\chi^2$ analysis.

\section{Discussion and conclusions}
\label{disc}

In this paper we studied the critical behaviour of the 2$d$ 
Ising model with quenched random site dilution via the 
EEA. Our purpose was twofold: first 
we wanted to address again the question about the reliability 
of the method; second, we extended the study started in Ref.\ 
\onlinecite{Ku94} to systems of larger sizes, also in view of 
the new results which have appeared in the literature since then, 
\cite{Aa+97,St+97,Ball+97,Rod+98,Sel+98} in order to allow perhaps 
a clearer discrimination between the two contradicting pictures 
of the system's critical behaviour that survived so far, the 
logarithmic corrections and the weak universality scenario.

\subsection{Reliability of the Method}

As the description of quenched disorder within our EEA is only 
approximate, we have to worry about how good it actually is.
Before gathering the different pieces of evidence accumulated
during our numerical study confirming the validity of our approach, 
we would like to address the question from a more general viewpoint.
In Ref.\ \onlinecite{Ku94} the correlation length $\xi_k$ describing 
the asymptotic decay of the correlation function $G_k (r_{ij})$ of 
the {\em disorder\/} degrees of freedom was introduced and studied.
Its finite-size estimates can be computed from the ratio of two 
TM eigenvalues in complete analogy to (\ref{csi1}) by replacing
$\gamma_2$ with $\tilde\gamma_{2}$, where $\tilde\gamma_{2}$ is the 
second eigenvalue of the symmetric block of the transfer matrix $\Gamma$.

As regards the analysis of this new correlation length, we have
both bad and good news. The bad news is that $\xi_k$ actually diverges at 
criticality in our different approximating systems. This escaped our 
attention in Ref.\ \onlinecite{Ku94}: 
while $\xi_k$ was observed never to exceed a few lattice spacings for 
the accessible system size, its size dependence at $T_c$ had not been 
monitored.
However, before concluding from this that our systems 
provide only a rather poor description of quenched disorder, we have also 
determined the behaviour of the ratio $R_L= \xi_{k,L}(T_c) / \xi_L(T_c)$ 
of these two correlation lengths at criticality, and the results of this 
study may be taken for the good news. We find that $R_L\to {1}/{8}$ 
at large $L$, independently of $\rho$. Since this limit represents nothing 
but the ratio $\eta/\eta_k$ of the anomalous dimensions of the spin and 
occupation-variable correlation functions at $T_c$, this implies that 
$\eta_k = 2$, hence unusually large. The $k_i$-correlator is thus 
`almost summable' at criticality (its sum  has a logarithmic infrared 
divergence). Note that the ratio $G_k(r)/G(r)$ behaves as
\be
\frac{G_k(r)}{G(r)} \sim \frac{1}{r^{7/4}}
\ee
and thus decays to zero, stating that $k_i$ correlations are negligibly 
small if compared to $\sigma_i$ correlations at large distances. 
This is why the system may be regarded as effectively quenched despite 
remaining correlations between the $k_i$, and it may be regarded as the
reason why our results compare so favourably with those obtained via 
more conventional approaches.

Indeed, both checks against exact results, wherever obtainable, 
and direct comparison of our data with those obtained by the random TM  
approach (which allows for an in principle exact treatment 
of the disorder) increase our confidence in the validity of our method 
beyond any reasonable doubt. 
Among the exact results correctly reproduced by our method we mention: 
{\em i)\/} data obtained for a one-dimensional system; \cite{Kunot} 
{\em ii)\/}  the value of the initial slope of the critical line in the phase
	diagram, reproduced to within 0.01\%;
{\em iii)\/} the correct value of the connectivity length exponent $\nu_p$ and 
	of the crossover exponent $\phi=\nu_p/\nu=1$ at percolation; \cite{Ku94}
{\em v)\/}  the precision with which the values of the exponent $\eta$, of the
           ratio $\gamma/\nu$, and of the central charge are determined
           for all the points investigated in the phase diagram.
In addition, as mentioned above, the random TM data (obtained in the case of 
bond disordered systems) \cite{Aa+97,St+97} exhibit the same finite-size 
signature, and qualitatively the same behaviour as regards their dependence on 
the disorder strength as those obtained in the present paper.
 
A definite advantage of our method over other numerical methods 
dealing with disordered systems, is that it does not suffer from
non-self-averaging difficulties, and provides directly through the 
ratio of TM eigenvalues the average correlation length, not the typical 
one, such that the successive analysis turns out to be more 
straightforward than, e.g., in Ref.\ \onlinecite{Aa+97}.

The results discussed so far were all obtained within approximating system 
{\em (d)\/}. Comparative studies including results also from the other systems
{\em (a)--(c)\/} were presented in Ref.\ \onlinecite{Ku94}. 
They showed that systems 
{\em (a)--(d)\/} appear to be in the same universality class as regards 
their critical behaviour. Non-universal quantities, however, such as 
the critical temperature itself, or the value of the percolation 
threshold $\rho_c$, do depend on the approximating system, differing 
between those with and without the plaquette constraint and 
those estimated by other numerical works. \cite{Ball+97,perc}
The conclusion is that the description of quenched disorder, though
only approximate, appears to be precise enough to put all our 
approximating systems into the same universality class as the the 
fully quenched system, and this is all that is required for the purpose 
of the present study. 

\subsection{Discriminating Between the Scenarios}

 We now come to our second point, concerning the discrimination
between the two scenarios proposed for the critical behaviour of the 
model considered in this paper. Our view is that some of the FSS 
investigations which have appeared in the literature -- including the 
recent ones \cite{TalShu94,Aa+96,Aa+97,St+97,Ball+97,Sel+98,KiPa94,Ku94} 
-- may perhaps not be as decisive on this matter as their authors have
tended to believe. We shall now discuss some issues on which, we 
believe, clarification or amendments are needed, presenting a critical
review of available material in the light of our own data analysis.

The results of Refs.\ \onlinecite{KiPa94} and \onlinecite{Ku94} have been 
interpreted as 
providing evidence in favour of a weak universality scenario. The former
is a Monte Carlo investigation both at, and off criticality of the 
site diluted system. On the basis of $\chi^2$ data analyses, Kim and 
Patrascioiu concluded that their off-critical simulations of susceptibility 
and correlation length were better described by modified power laws 
than in terms of the log-corrections scenario. No analogous discrimination
was attempted for the specific heat, and hence no check of hyperscaling
was performed. The main weakness of their 
off-critical simulations is that reduced temperatures are still sizeable, 
and the constancy of the ratio $\gamma/\nu$ appears slightly less well 
satisfied than the individual errors on the values of $\gamma$ and $\nu$ 
would allow. Simulations at criticality yielded a specific heat very slowly
increasing with system size, and arguments in favour of a saturation
were advanced. The constancy of $\gamma/\nu$ was clearly shown and the 
validity of the Fisher relation confirmed, but -- as noted above -- this 
cannot be taken to support either scenario over the other.

Our own earlier TM strip-scaling results,\cite{Ku94} too, were interpreted in 
terms of weak universality. Possible effects of logarithmic corrections in 
the effective size-dependent exponents $\nu_L$ were looked for, but were either
absent or simply not discernible due to the rather moderate strip widths 
available in that study. Indeed, the rather limited system sizes may be taken 
as one of the weakest points of that investigation. Moreover, specific heats
had not been computed and so hyperscaling within a varying exponents picture
was not checked. Again, the observed constancy of $\gamma/\nu$ and $\eta$ may
be regarded as a piece of evidence for the reliability of the method, but not
in favour of either scenario, except in so far as implying, that {\em if\/} 
critical behaviour in the model were non-universal, the observed 
non-universality would have to be of the weak form.

Let us now turn to the recent investigations which have been taken to support 
the log corrections scenario. They are either of the TM strip-scaling 
\cite{Aa+96,Aa+97,St+97} or of the Monte Carlo \cite{TalShu94,Ball+97,Sel+98} 
type. 

The strip scaling data of Aar\~{a}o Reis {\em et al.\/} \cite{Aa+96,Aa+97} 
for the correlation length in the bond disordered system appeared to give 
values of $\nu$ slightly greater than that of the pure system, which the 
authors, however, discarded, attributing them to preasymptotic effects 
originating from logarithmic corrections. Their interpretation of the data 
in these terms does, indeed, describe the data rather well on the level of
the $\mu_L$, but {\em not\/} on the level of the size dependent exponents
$\nu_L$. 
Moreover, we recall our discussion of this point in 
Sec. \ref{resu} D, according to which the power law picture provides a 
consistent interpretation of the data as well.

Turning to the critical specific heat data presented in Refs.\
\onlinecite{Aa+97} and \onlinecite{St+97}, they were interpreted as 
``clearly suggesting a divergence in the thermodynamic limit". This 
point was claimed to be strengthened by pushing 
system size up to $L=18$
(and in less precise simulations even to impressive $L=23$), and by noting 
that specific heat data 
appeared to be halfway between single- ($\ln L$) and double-
($\ln \ln L$) logarithmic dependence on strip width, 
from which a divergence 
in the thermodynamic limit was inferred. One should, however, note that 
at moderate $L$ this kind of behaviour is to be expected in {\em both\/} 
scenarios if the correlation length exponent is close to 1 and, hence, 
$\alpha$ only slightly negative. The difference 
is that $C_L$ will continue to increase 
at least as fast as $\ln\ln L$ for 
{\em all\/} $L$, in the log-corrections scenario, whereas there 
will be a crossover 
to a growth which is slower than $\ln \ln L$  
at $L_* \simeq \exp\{ -\nu/\alpha \}$ in case of 
a modified power law (\ref{cvofl2}) with $\alpha < 0$. Tentatively 
accepting the value $\nu\simeq 1.083$ determined by Aar\~{a}o Reis 
{\em et al.\/} \cite{Aa+96} for the $r\equiv J_1/J_2 = 0.25$ case 
investigated in Ref.\ \onlinecite{St+97}, one would have to locate the 
crossover length at $L_* \simeq 680$. Thus $L_{max}=23$ is still 
much too small to allow concluding in favour of either scenario.

We have carried out $\chi^2$ data analyses on the raw data
of Ref.\ \onlinecite{St+97}, comparing the two scenarios, as we did on 
our own data. Power law fits and fits according to (\ref{cvofl1}) show 
no significantly different quality. In particular, it seems that a double
logarithmic law provides a better fit to the data selected from a window 
$[L_{\rm min},23]$ with $L_{\rm min}$ up to 8, while power laws give a 
smaller $\chi^2$ for the larger 
$L_{\rm min}$ up to $13$; for still larger $L$ the 
result of the fits become questionable for both hypotheses. Other kinds 
of fits have been tried, e.g. by selecting a movable window 
$[L_{\rm min},L_{\rm min}+7]$ and letting $L_{\rm min}$ run over the data, 
but without any significant improvement. Only by discarding the larger $L$ 
values, which are still much too noisy (as it is obvious by a quick look 
at the derivative information plotted in Fig. 3 of Ref.\ \onlinecite{St+97}), 
one gets values of $\chi^2$ slightly in favour of the double logarithmic 
form (\ref{cvofl1}). Incidentally, however, the power-law fits 
lead to an estimate for $\alpha/\nu$ compatible via hyperscaling with the 
$\nu$ value reported in Ref.\ \onlinecite{Aa+96}. This analysis cannot 
thus be seen as conclusive.

The Monte Carlo study of site diluted system by Ballesteros {\em et 
al.\/} \cite{Ball+97} shows critical specific heat data reasonably well 
fitted (in terms of $\chi^2$) by a double logarithmic form, at least 
for intermediate dilution. The same is true for a recent study of Selke 
{\em et al.\/}\cite{Sel+98} However, no power-law fitting is attempted for 
comparison. In earlier work on bond disordered systems,\cite{An+90} such
power-law fits had been attempted, and were regarded inferior to double
logarithmic ones. Note, however, that the maximum system sizes 
studied in Ref.\ \onlinecite{An+90} are for $r=0.25$ still below, and for 
$r=0.1$ above but still very close on a double logarithmic scale to the 
crossover lengths $L_*$ expected from the $\nu$ values reported in Ref.\ 
\onlinecite{Aa+96}.

The $\nu_L$ data presented in Ref.\ \onlinecite{Ball+97} have, in our
view, somewhat too large error bars to allow concluding with confidence  
that $\nu_L \to 1$ for large $L$ for all the values of $\rho$
also because Ballesteros {\em et al.\/}
{\em force\/} the intercept through 1 rather than fitting
it. The reasonably large values of $L$ reached in this investigation would
probably set the system in the asymptotic regime (at least for the stronger
disorder), and the FSS expression used by Ballesteros {\em et al.\/} is indeed
$\nu_L = 1 + A'/\ln L$. Nonetheless the constant $A'$ in this expression 
should, strictly speaking, come out to be 1/2, {\em independently\/} of the 
degree of disorder (see Eq.\ \ref{asynul}), a property which their data and their 
fits do {\em not\/} respect. 

In the off-critical Monte Carlo study of Talapov and Shchur\cite{TalShu94}, 
bond disorder was observed to lead to increased values for the magnetization 
and susceptibility exponents $\beta$ and $\gamma$, while the behaviour of the 
specific heat showed good agreement with the double logarithmic form 
(\ref{cvlog}). Together with the Rushbrooke equality $\alpha + 2\beta + \gamma
=2$, these findings embody an inconsistency from which the authors concluded
that their increased values for $\beta$ and $\gamma$ cannot be asymptotic.
On the other hand, it turns out that the specific heat data of Ref.\
\onlinecite{TalShu94} can be fitted equally well to a power law form with
a value of $\alpha$ compatible via the Rushbrooke relation with the increased
values of $\beta$ and $\gamma$.\cite{Tpriv} Note also that 
the ratio $\beta/\gamma$ in the disordered system is the same (to within 
a fraction of a percent) as in the pure system, as would be expected in a 
weak universality scenario, whereas individually $\beta$ and $\gamma$ 
change in the 6\% range.

The difficulties of FSS are avoided in the series expansion
study of Roder {\em et al.\/}; \cite{Rod+98} 
their determination 
of the exponent ${\tilde{\gamma}}$ in Eq.\ (\ref{chilog}), which appears 
to saturate at $7/8$ for sufficiently strong disorder,
may perhaps be taken as the strongest piece of evidence currently 
available in favour of the validity of the logarithmic corrections
scenario. Still, 
the analysis
suffers from the relatively small length of the series, 
at least in some regions of the phase diagram, and  
one would wish this to be made more
conclusive by including higher order terms, so as to 
reduce error bars.
It is also worth noting at this point that their claim that
$\chi_L(T_c)$ is unaffected by logarithmic corrections
cannot be upheld: see Eq.\ \ref{defchiL}.

We now turn to the present investigation. 
First, by combining finite size 
signatures of correlation length and domain wall free energy, we have in 
particular been able to locate  critical temperatures with extreme precision.
Second, we have significantly enlarged our system sizes. In interpreting 
our data, maximum care was constantly taken to be open in both directions. 
Based on results of Cardy and Ludwig, a more systematic FSS analysis than in 
previous numerical studies has been performed.  As we will presently show,
neither way of looking at the available 
finite size data is completely satisfactory.

The value of the effective central charge is found to be $c^{\prime}=\frac{1}
{2}$ with extremely high precision. If conformal invariance and reflection
positivity would {\em also\/} hold for disordered systems, this finding
would put the model into the Ising universality class and exclude continuously
varying exponents.\cite{conf} However, the results of Ludwig and 
Cardy\cite{Lud-Car} imply that the latter condition does {\em not\/} hold,
at least for the weakly disordered ferromagnetic Ising model.

Considering the correlation length exponent $\nu$, we observe that our
FSS estimates converge to values continuously varying with $\rho$. 
One might suspect, of course, 
that our extrapolations are misled 
because the algorithm would not pick up slowly 
varying logarithmic corrections. 
However, if we adopt this hypothesis,
we find it not easy to reconcile with the fact that not even the slightest
such effect is detectable in our extrapolations of $\eta$, $\gamma/\nu$ and
the central charge $c$, even though similar, albeit smaller 
offsets must
be expected to occur in these quantities as well, as is borne out by our
analysis of FSS 
in the Appendix.

The fact that the critical specific heat data may be well fitted both to  a
double logarithmic behaviour (\ref{cvofl1}) and to a cusp singularity 
(\ref{cvofl2}) shows that it is difficult to discriminate with confidence
between the two laws. However, we recall that a fit according to a modified 
power law (\ref{cvofl2}) entails a weak violation of hyperscaling, though
still within error bars.

In conclusion, as regards the two issues raised in this paper, 
we believe to have provided, if not a proof, satisfying
arguments in favour of the trustworthiness and accuracy of the
EEA to disordered systems. As to the second issue, 
we feel that further investigations would still be 
needed to provide a clearer and definite  
evidence in favour of either picture. The problems seen in the FSS analyses
of the system are not exclusive to our results, but are -- as we 
believe to have demonstrated -- common to virtually all previous FSS data
obtained so far.

Let us finally emphasize that we have no reasons to {\em a priori\/} distrust
the correctness of the theoretical picture that began to emerge through 
the work of Dotsenko and Dotsenko \cite{DD83} and the improved and corrected
versions of Shalaev \cite{Sha84}, Shankar \cite{Shan87} and Ludwig \cite{Lu88}. 
Still, in an ideal world, one would like to see this picture supported by 
numerical evidence much better than currently available.

\acknowledgements

We would like to thank Sergio de Queiroz for a nice interchange
of opinions and ideas, and John Cardy for illuminating correspondence. 
We are indebted to Dietrich Stauffer for kindly sending us the original 
data underlying the analysis of Ref. \onlinecite{St+97}. 
Conversations with Andreas Honecker, Giancarlo Jug,
Harald Kinzelbach, 
Robin Stinchcombe and Franz Wegner and are much 
appreciated too. G. M. gratefully acknowledges the Alexander von Humboldt- 
Stiftung for supporting his work and stay in Germany.

\appendix
\section*{FSS with Logarithmic Corrections}

In the present Appendix we collect the main results of the 
renormalization group (RG) analysis for the case where small amounts 
of disorder constitute a marginally irrelevant perturbation at the 
pure system's fixed point, thus giving rise to logarithmic corrections, 
as in the field theoretical approach to the {\em weakly disordered\/} 
2$d$ Ising model. 
We do this both for the sake of completeness, and to obtain sound 
understanding of the effects such corrections would have 
in a FSS analysis. This will turn out to be relevant not only in the 
asymptotic regime $L \gg 1$ but especially in the {\em preasymptotic\/} 
regime as specified below. 

We feel it particularly important to collect these results here, because the 
literature in the field abounds in heuristic derivations which have 
sometimes produced erroneous results and still more frequently 
even misconceptions as to what the effect of logarithmic corrections might 
be in the FSS signature of various thermodynamic functions and critical 
exponents. 

The following is basically exploring results Ludwig and Cardy \cite{Lud-Car} 
obtained in a replica approach to the problem, which are in turn based on 
earlier results on logarithmic corrections obtained by Cardy in a more general 
context.\cite{Car86}

If $g$ denotes the marginally irrelevant coupling arising from disorder
averaging in the (replicated) Hamiltonian, and by $\{u_n\}$ the set
of other scaling fields at the pure systems critical point, the RG 
flow-equations read
\bea
\frac{d g}{d l} & = & - \pi b g^2 + \cO(g^3)\\
\frac{d u_n}{d l} & = & y_n u_n -2\pi b_n g u_n + \cO(g^2 u_n)\ ,
\eea
where the $y_n$ are the eigenvalues of the linearized RG equations and 
the $b_n$ are operator product expansion (OPE) coefficients, which can be
formulated in terms of corresponding three-point functions. \cite{Lud-Car} 

Up to the order shown, these equations integrate to
\be
g(l) = \frac{g_0}{1 + \pi b g_0 l}
\label{gl}
\ee
with $g_0 = g(0)$ and
\be
u_n(l) = u_n(0) {\rm e}^{y_n l} (1 + \pi b g_0 l)^{-2b_n/b} \equiv 
u_n(0) \tilde u_n(l)\ .
\label{unl}
\ee

Among the $\{u_n\}$, the two relevant fields are $u_\varepsilon$ which
couples to the energy density and has RG eigenvalue $y_\varepsilon = 1$, 
and $u_\sigma$ coupling to the magnetization density with $y_\sigma= 15/8$. 
One may thus put $u_\varepsilon(0) = t$ and $u_\sigma(0) = h$. In a finite 
system of linear extent $L$ one may formally interpret $L^{-1}$ 
as another relevant scaling field with RG eigenvalue 1.

In the context of the replica formulation, the OPE coefficients  and the 
coupling $g$ are normalized such that the zero--replica limit of the product
$b g_0$ and of the ratios $b_n/b$ exist and are finite. The results of 
Ludwig and Cardy imply $b g_0 = 8 \Delta$ and $b_\varepsilon/b =1/4$ in
this limit, whereas $b_\sigma/b = 0$. Here $\Delta$ denotes the (bare) disorder
strength: $\Delta \propto \rho (1 -\rho)$ for a randomly site diluted model

The RG equations for the singular part of the free energy and the correlation 
length read
\be
f(t,h,g_0,L^{-1}) = {\rm e}^{-dl} f(u_\varepsilon(l),u_\sigma(l),g(l),
{\rm e}^lL^{-1})
\label{fl}
\ee
and
\be
\xi(t,h,g_0,L^{-1}) = {\rm e}^l \xi (u_\varepsilon(l),u_\sigma(l),g(l),
{\rm e}^lL^{-1})\ .
\label{xil}
\ee
Irrelevant scaling fields providing additional corrections to scaling have 
for simplicity been suppressed in these expressions. As usual, critical 
behaviour characteristic of the infinite system is obtained by considering $f$
and $\xi$ and their derivatives with respect to temperature (and field) at a 
scale $l$ chosen such that $u_\varepsilon(l) = \pm 1$. 
For $|t| \ll 1$ this gives
\be
{\rm e}^{y_\varepsilon l} = \frac{1}{|t|} \la 1 + \frac{\pi bg_0}
{y_\varepsilon}\ln\frac{1}{|t|} \ra^{2 b_\varepsilon/b}\ .
\ee
Inserting the values for the OPE coefficients reported in Ref.\ 
\onlinecite{Lud-Car}, and $\nu=1/y_\varepsilon= 1$ for the pure 2$d$ 
Ising model, this gives the divergence of the correlation length at $h=0$ 
to leading order
\be
\xi(t) \sim |t|^{-1} \la 1 + 8 \pi \Delta \ln\frac{1}{|t|}\ra^{1/2}\ .
\ee
The fact that $b_\sigma/b = 0$ entails that magnetization and susceptibility
will {\em to leading order\/} scale as pure powers of the correlation length, 
in agreement with earlier results of Shalaev\cite{Sha84}.

Phenomenological renormalization is based on analyzing the finite size
signatures of (\ref{xil}) and of its temperature derivative

\bea
\mu_L(t) & \equiv & \frac{d}{d t}\ \xi(t,h,g_0,L^{-1}) \nonumber \\
& = & {\rm e}^l\ \tilde u_\varepsilon(l)\ \xi_\varepsilon (u_\varepsilon(l),
u_\sigma(l),g(l),{\rm e}^lL^{-1})\ ,
\label{dtxil}
\eea
where $\xi_\varepsilon$ designates the partial derivative of the 
correlation-length scaling function (\ref{xil}) with respect to 
$u_\varepsilon(l)$. Finite size signatures at criticality ($t=0$, $h=0$)
are obtained by analyzing these quantities at a scale ${\rm e}^l = L$, 
giving
\bea
\xi^{-1}_L & = & L^{-1}\ \xi^{-1} (0,0,g(\ln L), 1)\nonumber\\
& = & L^{-1}\, \Phi(g(\ln L))
\label{defxiL}
\eea
and
\bea
\mu_L & =&  L^2  (1 + 8\pi \Delta \ln L)^{- 1/2}\, \xi_\varepsilon (0,0,g(\ln 
L), 1) \nonumber \\
& = & L^2  (1 + 8\pi \Delta \ln L)^{- 1/2}\, \Phi_\varepsilon (g(\ln L))
\label{defmuL}
\eea
respectively. Eqs.\ (\ref{defxiL}) and (\ref{defmuL}) 
define the universal scaling functions $\Phi(x)$ and 
$\Phi_\varepsilon(x)$ of the argument $x$ which, in the zero-replica 
limit, is $x= \Delta/(1 + 8 \pi\Delta \ln L)$. In the weak disorder limit or
at large $L$ this quantity is small, and one assumes that an expansion
of these scaling functions in $x$ exists. For $\Phi$ one gets $\Phi(x) =
\phi_0 + \phi_1 x + \phi_2 x^2 + \cO(x^3)$, the coefficients being known to 
be $\phi_0=\pi \eta$ from conformal invariance,\cite{Cardy} 
and $\phi_1 =0$ on account of the vanishing of $b_\sigma$. Thus 
\bea
\xi^{-1}_L & = & L^{-1}\,\Big[ \pi \eta + \phi_2 \Big( \Big(\frac{\Delta}{1+ 
8\pi \Delta \ln L}\Big)^2\Big)\nonumber\\
& & + \cO \Big(\Big(\frac{\Delta}{1+ 8\pi \Delta \ln L}\Big)^3\Big) \Big]\ .
\label{xiL}
\eea
A similar expansion is expected to hold for $\Phi_\varepsilon$, i.e.,
$\Phi_\varepsilon(x) = \phi_{\varepsilon0} + \phi_{\varepsilon1} x + 
\cO(x^2)$, so (with $D_1= \phi_{\varepsilon1}/\phi_{\varepsilon0}$)
\bea
\mu_L & = & \phi_{\varepsilon0} L^2 \la 1 + 8\pi \Delta \ln L\ra^{- 1/2} 
\times \Big[ 1 + \frac{D_1 \Delta}{1+ 8\pi \Delta \ln L} \nonumber \\
& & \quad +\cO \Big(\Big(\frac{\Delta}{1+ 8\pi \Delta \ln L}\Big)^2\Big) \Big]
\label{muL}
\eea
though the coefficients $\phi_{\varepsilon0}$ and $\phi_{\varepsilon1}$ are to 
the best of our knowledge not known. Interest in this quantity derives from 
the fact that, within the phenomenological renormalization group scheme, the 
correlation length exponent $\nu$ is obtained by computing  the sequence 
of finite size approximants $\nu_L$, defined by
\be
\nu_L^{-1} = \frac{d \ln  \mu_L }{d \ln L} - 1\ ,
\label{defnuL}
\ee
and by extrapolating this sequence to the large system limit. Eq. (\ref{nuofl})
is just the finite difference approximation to (\ref{defnuL}).

Similarly, the FSS analysis for the susceptibility gives
\bea
\chi_L & = & L^{\gamma/\nu}\ f_{\sigma\sigma}(0,0,g(\ln L),1)
\nonumber \\
& = & L^{\gamma/\nu}\ \Phi_\chi (g(\ln L))
\label{defchiL}
\eea
with the same conventions for the notation. The ratio $\gamma/\nu$ is 
that of the pure system, i.e., $\gamma/\nu=7/4$. On account of the 
vanishing of $b_\sigma$, no additional logarithmic terms above those 
following from the expansion of $\Phi_\chi$ appear in (\ref{defchiL}), 
in contrast to what happens for $\mu_L$. As before, one assumes that an 
expansion of the form $\Phi_\chi(x) = \phi_{\chi0} + \phi_{\chi1} x + 
\phi_{\chi2} x^2 + \cO(x^3)$ exists, and defines the effective size 
dependent ratio $(\gamma/\nu)_L$ by $(\gamma/\nu)_L = \partial \ln 
\chi_L/\partial \ln L$. It will be useful to introduce the abbreviation 
$E_i=\phi_{\chi i}/ \phi_{\chi0}$ below.

For the interpretation of numerical data it is relevant to note results 
for these quantities both in the preasymptotic regime $8\pi\Delta \ln L \ll 1$ 
and in the asymptotic regime $8\pi\Delta \ln L \gg 1$.

(i) In the preasymptotic regime, they read 
\begin{mathletters}
\label{logpre}
\bea
\xi^{-1}_L & = &L^{-1}\,\la \pi \eta + \phi_2 \Delta^2 
                  (1- 16\pi \Delta \ln L) \ra  \label{precsi} \\
\mu_L & = & \phi_{\varepsilon0} L^2 \big( 1 +  D_1 \Delta - (1+ 2 D_1 \Delta) 
                        4\pi \Delta \ln L \big) \label{premul} \\
\nu_L & = & 1 + 4\pi \Delta + (4\pi \Delta)^2 \big( 1 + D_1/2\pi  - 
                        2 \ln L \big)          \label{prenul} \\
\Big(\frac{\gamma}{\nu}\Big)_L & = & \frac{7}{4} - 8 \pi\Delta^2 E_1 
     + 8\pi\Delta^3 \Big(E_1^2 - 2 E_2 \nonumber\\
      & & \qquad + 16 \pi E_1 \ln L \Big)             \label{pregam}
\eea
\end{mathletters}
where we have kept the lowest order in $\Delta \ln L$. Additional terms down 
by further factors of $\Delta$ or $\Delta \ln L$ are not shown. 
These results exhibit non-universal corrections.

(ii) In the asymptotic regime, on the other hand, one has
\begin{mathletters}
\label{logas}
\bea
\xi^{-1}_L & = & L^{-1}\,\Big[\pi \eta + \phi_2\Big(\frac{1}{8 \pi \ln L} 
          \Big)^2 \nonumber\\
& &\qquad + \cO \Big(\Big(\frac{1}{\ln L}\Big)^3 \Big)\Big] \label{asycsi} \\
\mu_L  & = & \phi_{\varepsilon0} L^2 \big(8\pi\Delta\ln L\big)^{-1/2} 
          \nonumber \\
& & \times \lb 1+ \frac{D_1} {8\pi \ln L} + \cO \Big(\Big(\frac{1}{\ln L}
          \Big)^2 \Big) \rb                           \label{asymul} \\
\nu_L & = & 1 + \frac{1}{2 \ln L} + \cO \Big(\Big(\frac{1}{\ln L}\Big)^2 
          \Big)\ ,                                    \label{asynul} \\
\Big(\frac{\gamma}{\nu}\Big)_L & = & \frac{7}{4} - E_1\frac{1}{8\pi (\ln L)^2} 
+\cO\Big(\Big(\frac{1}{\ln L}\Big)^3 \Big)\ ,         \label{asygam}
\eea
\end{mathletters}
so the asymptotic corrections to $\xi_L^{-1}$, $\nu_L$ and $(\gamma/\nu)_L$ 
turn out to be universal.

Ludwig and Cardy have also reported corresponding FSS expressions for the 
central charge, in both regimes.\cite{Lud-Car}. These are
\begin{mathletters}
\label{cL}
\be
c  =  \frac{1}{2} - 128 \pi^3\Delta^3 (1 - 24  \pi \Delta \ln L)
\label{precc}
\ee
and 
\be
c  =  \frac{1}{2} - \frac{1}{4} (\ln L)^{-3}\ + \cO \la\frac{1}{8 \pi 
      \ln L} \ra^4 \ ,
\label{asycc}
\ee
\end{mathletters}
respectively, where we have once more kept the lowest order in $\Delta \ln L$ 
in the preasymptotic regime. 

Additional terms further down by powers of $L^{y_\alpha} (1 + 8\pi \Delta \ln 
L)^{-2 b_\alpha/b}$, with $y_\alpha \lesssim -2$, will appear due to irrelevant scaling fields.

Eqs.\ (\ref{logpre}) - (\ref{cL}) are of great importance when analyzing FSS 
data, if logarithmic corrections are expected to be present. Especially in a 
strip-scaling approach such as ours, the preasymptotic results may turn 
out of particular relevance, since the maximum size available is such that 
one may not reach the asymptotic regime, which is likely to be true at 
least for weak disorder. This has, indeed, been the point of view 
emphasized by Aar\~{a}o Reis et al.\cite{Aa+97}
\bigskip
\hfil\hrule\hfil

\end{document}